\documentclass[prl,aps,twocolumn,floats,showpacspsfig]{revtex4}
\usepackage{amssymb}

\usepackage{epsfig}

\newcommand{\beq}{\begin{equation}}
\newcommand{\eeq}{\end{equation}}
\newcommand{\bea}{\begin{eqnarray}}
\newcommand{\eea}{\end{eqnarray}}

\begin{document}
\title{Spin-liquid model of  the sharp resistivity drop in $La_{1.85}Ba_{0.125}CuO_4$.}

\author{A. V. Chubukov$^{1}$ and A.  M. Tsvelik$^{2}$}
\affiliation{$^{1}$ Department of University of Wisconsin, Madison, WI 53706, USA}
\affiliation{$^{2}$
Department of  Condensed Matter Physics and Materials Science, Brookhaven National Laboratory, Upton, NY 11973-5000, USA}
\date{\today}

\begin{abstract}
We use the phenomenological model proposed in our previous paper [Phys. Rev. Lett. {\bf 98}, 237001 (2007)] to analyse the magnetic field dependence of the onset temperature for two-dimensional fluctuating superconductivity $T^{**} (H)$. 
We demonstrate that the slope of $T^{**} (H)$ progressively goes down as $H$ increases, such that the upper critical field progressively increases as $T$ decreases. The quantitative agreement  with the recent measurements of $T^{**} (H)$ in $La_{1.85}Ba_{0.125}CuO_4$ is achieved for  the same parameter value as was derived in our previous publication from the  analysis of the electron self energy. 
\end{abstract}

\maketitle

Recent experiments on $La_{1-x}Ba_{x}CuO_4$ at $x=1/8$ ~\cite{li} revealed a complex hierarchy of energy scales in this material.  It displays a charge ordering transition at $T_{co} = 54K$, a spin ordering transition at $T_{spin} = 42K$ with a subsequent one order of magnitude  drop  in the in-plane resistivity, the Berezinskii-Kosterlitz-Thouless (BKT) transition to a two-dimensional superconductivity at $T_{BKT} = 16K$, a crossover from 2D to 3D regime around $10K$, and a transition to a true 3D superconductivity at $4K$. This hierarchy is summarized and discussed in detail in ~\cite{berg}.

 It turns out that the temperature $T^{**}$ where the resistivity crossover occurs is sensitive to the $c$-axis magnetic field which separates  this phenomenon separately from the spin ordering. In this paper, we address the issue of this  crossover.  The  measurements 
 performed in a magnetic field ~\cite{li} revealed that (i)
 $T^{**}$  marks the onset of fluctuational diamagnetism, and  (ii) 
$T^{**}$  decreases with the field.
These  two effects and the fact that the resistivity 
 sharply drops  $T^{**}$ are  consistent with the idea 
 that $T^{**}$ marks the onset of a
 fluctuational pairing regime without (quasi-) long-range superconducting order. The details of the system behavior near  $T^{**}$, however, depend on the underlying model. The authors of ~\cite{berg} considered a model of 
 weakly coupled parallel superconducting stripes.  
Within this model,  $T^{**}$ is the temperature at which
 the inter-stripe coupling becomes strong, and a vortex liquid is formed. 

We propose another explanation, 
 based on the model with a flat Fermi surface
 in the antinodal regions near 
$(0,\pi)$ and $(\pi,0)$ points in the Brillouin zone~\cite{tsv_ch}. 
 Fermions in these regions form two quasi-1D spin liquids  coupled by 
Josephson-type interaction.  
In this model, the pairing amplitudes in the antinodal regions are developed at $T^* \gg T^{**}$ due to the
 attractive  interactions in the spin-liquid state, however, 
 phase fluctuations at $T >> T^{**}$ are effectively one-dimensional, 
and are pinned by the defects.  At $T^{**}$, the Josephson coupling becomes 
sufficiently strong  to lock the relative phase of the two order parameters at $\pi$, 
and the system response becomes two-dimensional. This leads to 
depinning of the phase fluctuations resulting in the drop in  the resistivity.
Still, because of vortices in the 2D regime, the (quasi)-long-range superconducting order develops only at a smaller $T_c < T^{**}$.  

\begin{figure}
\begin{center}
\epsfxsize=0.35\textwidth
\epsfbox{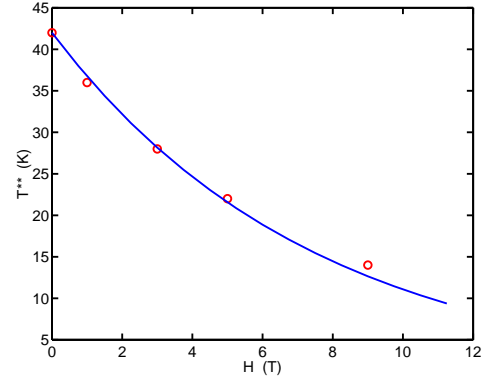}
\end{center}
\caption{The plot of $T^{**}(H)$. The points are the data
 from Ref.~\cite{li}, the solid curve is the exponential fit by Eq. 
(\ref{1}). }
\label{net}
\end{figure}      

Just like the model of parallel stripes\cite{berg},
 our model of ``crossed stripes'' near $(0,\pi)$ and $(\pi,0)$ 
explains qualitatively  the resistivity drop, the absence of fluctuational diamagnetism above $T^{**}$, and the sensitivity of $T^{**}$ to a magnetic field.~\cite{comm}.
 However, the measurements of $T^{**} (H)$  put an additional
 constraint on 
 the theory -- not only $T^{**}$ decreases with the field, but $|d T^{**}/dH|$ 
also {\it decreases} as $H$ goes up, i.e., at very low $T$, the critical field 
below which the system response is two-dimensional, becomes very large.
  The data for $H < 9T$ can be well  fitted by the exponential dependence (see Fig. 1):
\beq      
\frac{T^{**} (H)}{T^{**} (0)} = exp\left(-H/H_0\right),  ~~H_0 \approx 7.5T 
\label{1}
\eeq
For such $T^{**} (H)$,  $|d T^{**}/dH|$ exponentially 
decreases as $H$ increases. If this trend continued to higher $H$, 
 the critical field  $H_{c2} (T)$ defined as $T^{**} (H_{c2}) = T$ would 
become infinite at $T=0$.

The $H$ dependence of $T^{**}$ for Josephson-coupled 
stripes running parallel to each other in the 2D plane, i.e., for the same 
model as in Ref.~\cite{berg} was considered by
 Carr and one of us~\cite{carr}.  It was  found that the 
 slope of   $d T^{**}/dH$ increases with decreasing $T$, and $H_{c2}$  remains finite at $T=0$, in qualitative disagreement with the data. 
We demonstrate below that our model of crossed stripes located
 near $(0,\pi)$ and $(\pi,0)$ yields the behavior of $T^{**} (H)$ in a good agreement  with the measurements. Thus we show that the slope of  $d T^{**}/dH$ decreases with increasing $H$ for any value of the scaling dimension $d$ of the superconducting 
 order parameter. To achive a quantitative agreement with the experimental fit (\ref{1}) we have to set  $d \approx 1/2$. We have to remind the reader that in \cite{tsv_ch} the same value of $d$ was postulated on the basis of analysis of the electron self energy. This gives an important check for self-consistency of the theory.  

We associate $T^{**} (H)$ with the instability of a 2D pairing susceptibility in the random phase approximation (RPA). Fluctuations beyond RPA transform the instability into a crossover~\cite{tsv_ch}.  In zero field, the  RPA expression 
 for the susceptibility reads, in momentum space
\begin{equation}
\chi (k_x, k_y)  = \chi_0 (k_x) + J^2 \chi (k_x, k_y)~ \chi_0 (k_x)~
 \chi_0 (k_y)
\label{2_11}
\end{equation}
where $\chi_0 (k)$ is the 1D static pairing susceptibility~\cite{schulz}:
\begin{widetext}
\beq
\chi_0 (k) = \frac{2}{\Delta^2} \left[ \sin{\pi d}~\Gamma^2\left(1-d\right) \left(\frac{2\pi T}{\Delta}\right)^{-2+2d} \left|\frac{\Gamma\left(d/2 + i v q/4\pi T\right)}{\Gamma\left(1-d/2 + i v q/4\pi T\right)}\right|^2 -\frac{\pi}{1-d}\right]
\label{3_11}
\eeq
\end{widetext}
Here $\Gamma (...)$ are $\Gamma-$functions, $d<1$ is the scaling dimension of the superconducting order parameter, $v$ is the velocity of the phase mode, and $\Delta$ is the ultraviolet cut-off. The last term in $\chi_0$ can be neglected as we will only consider $T \ll \Delta$, when the first term in (\ref{3_11}) dominates.
 Parameters $v$ and $d$ are free parameters of our theory and should be extracted from the experiments in the $T$ region  where the superconducting phase fluctuations are essentially one-dimensional (that is, at $T$ 
below the spin gap, but larger
 than $T^{**}$). In \cite{tsv_ch} we found that the best agreement with the photoemission experiments is obtained when  $d \approx 1/2$. As we will see, this value is also favored  by the observed $T^{**}(H)$ dependence. 

Taking a Fourier transform over $k_x$, but leaving $k_y$ intact, 
 we obtain from  (\ref{2_11}):
\begin{eqnarray}
&& \chi_{k_y} (x- x_1) = \label{2}\\
&& \chi_0 (x-x_1) + J^2\int d x^\prime \chi_0 (k_y)\chi_0 (x-x^\prime) \chi_{k_y} (x^\prime - x_1)\nonumber
\end{eqnarray}
In a magnetic field, $k_y \rightarrow k_y + H x^\prime$ (we set $2e/c =1$). 
Setting $k_y =0$ and $x_1 =0$, we obtain integral equation for $\chi (x) = \chi_{k_y =0} (x)$ in the form
\begin{equation}
\chi (x) = \chi_0 (x) + J^2 \int d x^\prime \chi_0 (x-x^\prime) \chi (x^\prime) \chi_0 (H x^\prime)
\label{2_1}
\end{equation} 
where $\chi_0 (H x^\prime)$ is given by (\ref{3_11}) for 
$k = H x^\prime$, and $\chi_0 (x)$ is the Fourier transform of 
$\chi_0 (k)$.  The temperature 
$T^{**} (H)$ is the one at which $\chi (x)$ diverges.\\ 

{\it Weak fields.}~~~~Consider first the case when the magnetic field is weak, i.e., $T^{**} (H) = T^{**} (0) (1 - \delta T)$, and $\delta T \ll 1$. 
 A simple analysis shows that the parametrical 
condition for a weak field is  $v^2 H/T <<1$.
  Expanding $\chi_0 (H x^\prime)$  in $H$, we obtain from (\ref{3_11})
\begin{equation}
\chi_0 (H x^\prime) = B_d \left(\frac{2\pi T}{\Delta}\right)^{2d-2}~\left[1 - A_d \left(\frac{v H x^\prime}{\pi T}\right)^2\right]
\label{3}
\end{equation}
where 
\bea
A_d &=& \frac{1}{16} \left[\psi^{(1)} \left(d/2\right) - \psi^{(1)} \left(1- d/2\right)\right],\nonumber \\
 ~~B_d &=& \frac{2}{\Delta^2} \sin{\pi d}~\Gamma^2\left(1-d\right) \frac{\Gamma^2\left(d/2\right)}{\Gamma^2\left(1-d/2\right)} 
\label{3_1}
\eea 
and $\psi^{(1)} (x)$ is the derivative of the diGamma function.

Substituting (\ref{3}) into (\ref{2_1}), we obtain an integral equation for $\chi (x)$ in the form
\begin{eqnarray}
&&\chi (x)  = \chi_0 (x) + J^2 \int d x^\prime \chi_0 (x-x^\prime) \chi (x^\prime) \chi_0 (0) \nonumber \\
&& - J^2 \chi_0 (0) A_d \frac{v^2 H^2}{(\pi T)^2} \int d x^\prime 
\chi_0 (x-x^\prime) \chi (x^\prime) (x^\prime)^2
\label{4}
\end{eqnarray}
where $\chi_0 (0) = \chi_0 (k=0)$. 
Taking Fourier transform back to momentum space  ($x \rightarrow k_x =k$), and integrating by parts, we re-write the integral equation for $\chi$ as 
\begin{eqnarray}
&&\chi (k) \left[1 - J^2 \chi_0 (k) \chi_0 (0)\right] - J^2 \chi_0 (k) \chi_0 (0)
~\frac{A_d v^2 H^2}{(\pi T)^2} \chi^{\prime \prime} (k)\nonumber \\
&& = \chi_0 (k)
\label{6}
\end{eqnarray}
This can be re-expressed as
\begin{equation}
\left(\epsilon + c_1 k^2 - c_2  \frac{\partial^2}{\partial k^2}\right) \chi (k) = \chi_0 (k)
\label{7}
\end{equation}
where $\epsilon = 1 - (T^{**}(0)/T)^{4-4d}$, $c_1 = A_d v^2/(\pi T)^2$, $c_2 = A_d v^2 H^2 /(\pi T)^2$, and we defined $T^{**} (0) = (\Delta/2\pi)~(B_d J)^{1/(2-2d)}$.
This agrees with the zero-field transition temperature in ~\cite{tsv_ch}.
Expanding now in the eigenvalues  of the differential equation as
\begin{equation}
\chi (k) = \sum_n a_n \chi_n (k),~~ \chi_0 (k) = \sum_n a^{(0)}_n \chi_n (k)
\label{8}
\end{equation}
where $\chi_n (k)$ are the solutions of
\begin{equation}
\left(c_1 k^2 - c_2  \frac{\partial^2}{\partial k^2}\right) \chi_n (k) = \epsilon_n \chi_n (k),
\label{9}
\end{equation}    
we obtain
\begin{equation}
a_n = \frac{a^{(0)}_n}{\epsilon + \epsilon_n}
\label{10}
\end{equation}
The eigenvalues of Eq. (\ref{9}) can be easily obtained as (\ref{9})
  can be re-expressed as a harmonic oscillator
\begin{equation}
- \frac{1}{2M} \frac{\partial^2 \chi_n (k)}{\partial k^2} + \frac{M \omega^2 k^2}{2} \chi_n (k)  = \epsilon_n \chi_n (k)
\label{11}
\end{equation}  
where $\omega^2 = 4 c_1 c_2$ and $M^{-1} = 2 A_d (v/\pi T)^2$.  
The eigenfunctions of (\ref{11}) are
 $\epsilon_n = \omega (n +1/2)$, the lowest one is $\epsilon_0 = \omega/2
 = A v^2 H/ (\pi T)^2$.
From (\ref{10}), the instability in the field occurs when
 $\epsilon + \epsilon_0 =0$, i.e, when $T = T^{**} (H) = 
T^{**} (0) (1 - \delta T)$, where
\begin{equation}
\delta T  \approx  \frac{1}{4(1-d)}~\frac{A_d v^2 H}{(\pi T^{**} (0))^2}
\label{12}
\end{equation}

We see that at small fields, $T^{**} (H)$ decreases linearly with $H$.
The linear dependence at small fields is also present in the model of parallel stripes~\cite{carr}. 
If we formally extrapolate the small-field result to $T=0$, we 
obtain the  upper critical field 
\beq
H^{extr}_{c2} (T=0) = \left(\frac{\Delta}{v}\right)^2 (J B_d)^{1/(1-d)} \frac{1-d}{A_d}
 \label{ex_1}
\eeq
\\
The actual $H_{c2} (T=0)$ is somewhat smaller in the model of parallel stripes~\cite{carr}, but, as we will see, is much larger than (\ref{ex_1}) in our model of crossed stripes. 

{\it Strong fields. }~~~~~ Consider now the opposite limit of vanishing $T$, when $v^2 H/T >>1$, i.e., the expansion in the field is no longer possible.
In this limit, we have from (\ref{3_11})
\begin{equation}
\chi_0 (H x^\prime) = \frac{{\bar B}_{d}}{|H x^\prime|^{2-2d}}
\label{14}
\end{equation}
where 
\bea
&&{\bar B}_{d} = (8/\Delta)^2 \sin (\pi d)~\Gamma^2 (1-d) (v^2/4\Delta^2)^d \nonumber \\
&&= B_d (2\Delta/v)^{2-2d} ~(\Gamma^2 (1-d/2)/\Gamma^2 (d/2)).
\eea
Instead of Eq. (\ref{6}), we now have
\begin{equation}
\chi (k) = \chi_0 (k) \left[1 + J^2 \frac{{\bar B}_{2d-1}}{H^{2-2d}} 
  \int d q \chi (q) \int d x^\prime \frac{e^{i (k-q) x^\prime}}{|x^\prime|^{2-2d}}\right]
\label{15}
\end{equation}
Using 
\begin{equation}
 \int d x^\prime \frac{e^{i (k-q) x^\prime}}{|x^\prime|^{2-2d}} = 
\frac{\Gamma (2d-1) \sin {\pi d}}{|k-q|^{2d-1}}
\label{16}
\end{equation}
and introducing 
\begin{equation}
\chi (k) = \frac{{\bar B}_{d}}{|k|^{2-2d}} {\tilde \chi} (k)
\label{17}
\end{equation}
and $d = (1 + \epsilon)/2$, we obtain from (\ref{15})
\begin{equation}
{\tilde \chi} (k) = 1 +  \frac{J^2 {\bar B}^2_d \cos (\pi \epsilon/2)~
 \Gamma (\epsilon)}{H^{1-\epsilon}} 
  \int d q \frac{{\tilde \chi} (q)}{|q|^{1-\epsilon} |k-q|^{\epsilon}}
\label{18}
\end{equation}
It is convenient to re-express this equation in the operator form, as 
 ${\hat L} {\tilde \chi} (k) = 1$, 
and expand in the eigenfunctions of the operator ${\hat L}$, which we label as 
${\tilde \chi}_m (k)$. We get 
\begin{equation}
{\tilde \chi} (k) = \sum_m \frac{a_m^{(0)}}{1 -\lambda_m}~{\tilde \chi}_m (k)
\label{19}
\end{equation} 
where $a_m^{(0)}$ are constants. The eigenvalues $\lambda_m$ are the solutions of 
\begin{equation}
{\hat L} {\tilde \chi}_m (k) = (1 -\lambda_m)  {\tilde \chi}_m (k)
\label{20}
\end{equation} 
where
\begin{equation}
{\hat L} {\tilde \chi}_m (k) = {\tilde \chi}_m (k) - 
 \frac{J^2 {\bar B}^2_d \cos {\pi \epsilon/2} \Gamma (\epsilon)}{H^{1-\epsilon}} 
  \int d q \frac{{\tilde \chi}_m (q)}{|q|^{1-\epsilon} |k-q|^{\epsilon}}
\label{21}
\end{equation} 

Eq. (\ref{21}) was studied in the context of non-BCS superconductivity
 (with frequency instead of momentum)~\cite{aace}. A similar equation has 
been  studied in the content of superconductivity in graphene~\cite{khvesh}.
 For $\epsilon >0$, the normalized solution of (\ref{21}) with the largest eigenvalue is 
\begin{equation}
{\tilde \chi}_m (k) = \frac{1}{|k|^\epsilon}
\label{22}
\end{equation} 
and the eigenvalue is 
\begin{equation}
\lambda_0  =  \frac{J^2 {\bar B}^2_d}{H^{1-\epsilon}} \Psi_\epsilon, 
~~ \Psi_\epsilon = \frac{\pi^2}{2} \frac{1}{\Gamma^2 (1-\epsilon/2) (\sin {\pi \epsilon/4})^2}
\label{23}
\end{equation} 

The critical field $H_{c2} (T=0)$ is determined from $\lambda_0 =1$ and is given by 
\beq
H_{c2} (T=0) = \left[J^2 {\bar B}^2_d  \Psi_\epsilon\right]^{1/(1-\epsilon)}
\label{25}
\eeq
In explicit form, we have
\begin{widetext}
\begin{eqnarray}
 && H_{c2} (T=0) = (J {\bar B}_d)^{1/(1-d)} \left(\frac{2\Delta}{v}\right)^2 ~
\left(\frac{8}{(2d-1)^2}\right)^{1/2(1-d)} ~\left[\frac{\Gamma\left(1-d/2\right)}{\Gamma\left(d/2\right)}\right]^{2/(1-d)} \nonumber \\
&&= H^{extr}_{c2} (T=0)  ~\left[\left(\frac{4 A_d}{1-d}\right)~\left(\frac{8}{(2d-1)^2}\right)^{1/2(1-d)}~~\left[\frac{\Gamma\left(1-d/2\right)}{\Gamma\left(d/2\right)}\right]^{2/(1-d)}\right]
\label{26}
\end{eqnarray}  
\end{widetext}

One can easily make sure that the actual $H_{c2} (T=0)$ is much larger than 
$ H^{extr}_{c2} (T=0)$ for all $d \leq 1/2$ for which our computational scheme is applicable. Furthermore, as $d$ approaches $1/2$, $H_{c2} (T=0)$ tends to infinity because $\Psi (\epsilon)$ diverges at vanishing $\epsilon = 2d-1$ 
as $\Psi_\epsilon \approx 8/\epsilon^2$.  The plot of the ratio $H_{c2} (T=0)/
H^{extr}_{c2} (T=0)$ is presented in Fig. 2.

\begin{figure}
\begin{center}
\epsfxsize=0.35\textwidth
\epsfbox{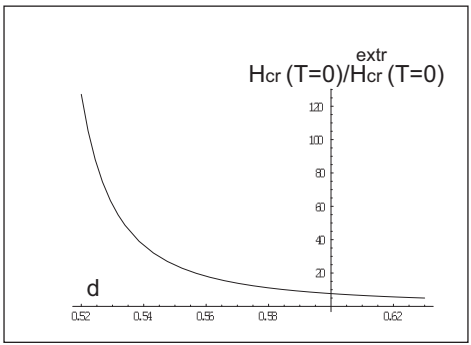}
\end{center}
\caption{The ratio $H_{c2} (T=0)/H^{extr}_{c2} (T=0)$ as a function of scaling dimension $d$, from Eq.(\ref{26}). The ratio diverges logarithmicaly 
at $d\rightarrow 0.5$.}
\end{figure}      

For $d \leq 1/2$, the analysis has to be modified to account for the divergence at $q=0$ in the r.h.s. of (\ref{21}). The expected result is that $H_{c2}$ becomes infinite at zero temperature. The divergence is power-law for $\epsilon <0$, and logarithmical at $\epsilon=0$. In the latter case,
\begin{equation}
\chi_0 (H x^\prime) = \frac{{\bar B}_{\epsilon =0}}{|H x^\prime|}
\label{14_1}
\end{equation}
and the RPA equation for $\chi (x)$ in the real space becomes
\beq
\chi (x) = -2 \log{T |x|} - \frac{2 (J {\bar B}_{\epsilon =0})^2}{H} \int \frac{dx^\prime}{|x^\prime |}~ \chi (x^\prime)~\log{(T|x-x^\prime|)}  
\label{27}
\eeq   
With the  logarithmic accuracy, we can approximate
\beq
\log{(|x-x^\prime|)} \approx \theta (x - x^\prime) \log{|x|} + \theta (x^\prime - x) \log{|x^\prime|}
\label{28}
\eeq   
Substituting into (\ref{27}), we re-write it as a differential equation
\beq
\partial^2_\zeta \chi + \frac{4 (J {\bar B}_{\epsilon =0})^2}{H} ~\chi = -2 \partial^2_\zeta \log{(T |e^\zeta -1|)}
\label{29}
\eeq
where $\zeta = \log |x|$.  The analysis of this equation shows that the susceptibility diverges at $H = H_{c2} (T) \propto |log T|$.  This is equivalent to 
$T^{**} (H) \propto exp{-H/H_0}$, in agreement with  Eq. (\ref{1}).  We see therefore that the 
 high field dependence is well captured by our model with $d \approx 1/2$ -- the same as we used in the previous work~\cite{tsv_ch} to fit the normal state self-energy. 
 
To summarize, we analyzed
 the behavior of $T^{**} (H)$  (or, equivalently $H_{c2} (T)$) in the model of two one-dimensional spin liquids near $(0,\pi)$ and $(\pi,0)$ coupled by
 Josephson-type interaction. For weak fields we found that
 $T^{**}$ decreases linearly with $H$. Extrapolating this dependence down to zero temperature yields
 the extrapolated field $H_{c2}^{extr}(T=0)$. 
 Considering the strong fields we found  that the actual $H_{c2} (T=0)$ is 
 always larger than the extrapolated value. The ratio 
$ H_{c2}(T=0)/H_{c2}^{extr}(T=0)$, characterizing the 
convexity of the $H_{c2}(T)$-curve, increases when $d$ decreases
 and becomes infinite at $d\leq 1/2$. This convex behavior is consistent with the data, and has to
 be contrasted with the {\it concave} behavior 
 for  the model of parallel stripes.  
As a further evidence in support of our model, we
 found that the experimental $H_{c2} (T)$  are well described by the theoretical formula with the scaling dimension 
of the 1D superconducting order parameter $d \approx 1/2$.
 The same $d$ provides the best fit to the photoemission data, as we argued
earlier~\cite{tsv_ch}. We think  that all these give our model a considerable advantage in treating  $La_{1.85}Ba_{0.125}CuO_4$.

We acknowledge useful discussions with E. Fradkin, S. Kivelson, D. Scalapino and J. Tranquada and to J. Tranquada for kindly providing us Fig. 1. The research was supported by  NSF-DMR 0604406 (A. V. Ch.), and by US-DOE under contact number DE-AC02-98 CH 10886 (A.M.T.).  AVC acknowleges the support from the Theory Institute for Strongly Correlated and Complex Systems at BNL.


\begin{thebibliography}{99}
\bibitem{comm}  The underlying physics behind our model is quite different from the one in the standard model of interacting stripes. From mathematical perspective, however, our model does indeed look like a model of intersecting 1D stripes, coupled by Josephson-like interaction.
\end{thebibliography}
\end{document}